\documentclass[a4paper]{jpconf}
\usepackage{graphicx}
     \usepackage{epsfig}

\newcommand{\GeV}{\textrm{GeV}}

\begin{document}

\begin{flushright}
LU TP 11-31\\
September 2011
\end{flushright}
%\vfill

\title{Diffractive Drell-Yan process: hard or soft?}

\author{Roman Pasechnik}

\address{Department of Astronomy and
Theoretical Physics, Lund University, SE 223-62 Lund, Sweden}

\ead{Roman.Pasechnik@thep.lu.se}

\author{Boris Kopeliovich}
 \address{ Departamento de F\'{\i}sica Universidad T\'ecnica
 Federico Santa Mar\'{\i}a; and\\
 Instituto de Estudios Avanzados en Ciencias e Ingenier\'{\i}a; and\\
 Centro Cient\'ifico-Tecnol\'ogico de Valpara\'iso;\\
 Casilla 110-V, Valpara\'iso, Chile}

\ead{Boris.Kopeliovich@usm.cl}

\begin{abstract}
Single diffractive Drell-Yan reaction in hadron-hadron collisions is
considered as an important source of information on
the properties of soft QCD interactions.
In particular, it provides an access to the
dynamics of the QCD factorisation breaking due to the interplay
between hard and soft interactions which leads to a nontrivial
energy and scale dependence of the Drell-Yan observables. We study
the process at forward rapidities in high energy proton-(anti)proton
collisions in the color dipole approach. Predictions for the total
and differential cross sections of the diffractive lepton pair
production are given at different energies.
\end{abstract}

\section{Introduction}

Diffractive large rapidity gap processes in QCD constitute a
noticeable fraction of all observed events. Depending on energy, it
may vary from a fraction of percent in $pp$ up to $10$ \% and more
in $ep$ scattering. Generally, such processes are very sensitive to
the soft and nonperturbative interactions despite the presence of a
hard scale, which makes them extremely difficult to investigate from
both the theoretical QCD and experimental points of view, and
intrinsic uncertainties of existing phenomenological models are
still quite large.

Diffractive Drell-Yan process at forward rapidities \cite{KPST06},
as well as diffractive heavy flavor production \cite{heavyF}, is one
of such processes, which give us an immediate access to soft QCD
evolution close to saturation regime. The understanding of the
mechanisms of inelastic diffraction came with the pioneering works
of Glauber \cite{Glauber}, Feinberg and Pomeranchuk \cite{FP56},
Good and Walker \cite{GW}. If the incoming plane wave contains
components interacting differently with the target, the outgoing
wave will have a different composition, i.e. besides elastic
scattering a new {\it diffractive} state will be created (for a
detailed review on QCD diffraction, see Ref.~\cite{KPSdiff}). In our
case, such a new state is given by the deeply virtual photon
radiation in the forward direction, which then can be seen as e.g.
heavy $\mu^+\mu^-$ pair in a detector.

The single-diffractive Drell-Yan reaction in $pp$ collisions is
characterized by a relatively small momentum transfer between the
colliding protons, such that one of them, e.g. $p_1$, radiates a
hard virtual photon $k^2=M^2\gg m_p^2$ and hadronizes into a
hadronic system $X$ both moving in forward direction and separated
by a large rapidity gap from the second proton $p_2$, which remains
intact, i.e.
\begin{eqnarray}
p_1+p_2\to \gamma^*(l^+l^-)+X+(gap)+p_2
\end{eqnarray}
Both the di-lepton and $X$ stay in the forward fragmentation region.
In this case, the virtual photon is predominantly emitted by the
valence quarks of the proton $p_1$. We will refer to this as the
diffractive Drell-Yan process at forward rapidities.

The dipole approach, previously applied to diffractive Drell-Yan
reaction in Ref.~\cite{KPST06}, led to the QCD factorisation
breaking, which manifests itself in specific features like a
significant damping of the cross section at high $\sqrt{s}$ compared
to the inclusive DY case. This is rather unusual, since a
diffractive cross section, which is proportional to the dipole cross
section squared, could be expected to rise with energy steeper than
the total inclusive cross section, like it occurs in the diffractive
DIS process. At the same time, the ratio of the DDY to DY cross
sections was found in Ref.~\cite{KPST06} to rise with the hard
scale, $M^2$. This is also in variance with diffraction in DIS,
which is associated with the soft interactions \cite{BP97}.

Long-range soft interactions between target and projectile
particles, or {\it the absorptive corrections} affect differently
the diagonal and off-diagonal terms in the hadronic current
\cite{PCAC}, in opposite directions, leading to an {\it unavoidable
breakdown of the QCD factorisation} in processes with off-diagonal
contributions only. Namely, the absorptive corrections enhance the
diagonal terms at larger $\sqrt{s}$, whereas they strongly suppress
the off-diagonal ones. In the diffractive DY process a new state,
the heavy lepton pair, is produced, hence, the whole process is of
entirely off-diagonal nature, whereas in the diffractive DIS
contains both diagonal and off-diagonal contributions
\cite{KPSdiff}.

Another reason of the QCD factorisation breaking is more specific
and concerns the interplay of soft and hard interactions in the DDY
amplitude. In particular, this leads to the leading twist nature of
the DDY process, whereas DDIS is of the higher twist \cite{KPST06}.
Large and small size projectile fluctuations contribute to the
diffractive DY process at the same footing, which further deepens
the dramatic breakdown of the QCD factorisation in DDY. This work is
devoted to a detailed study of consequences of such a breakdown in
typical DY observables.

\section{Diffractive Drell-Yan in dipole-target scattering}

\begin{figure}[h!]
\centerline{\epsfig{file=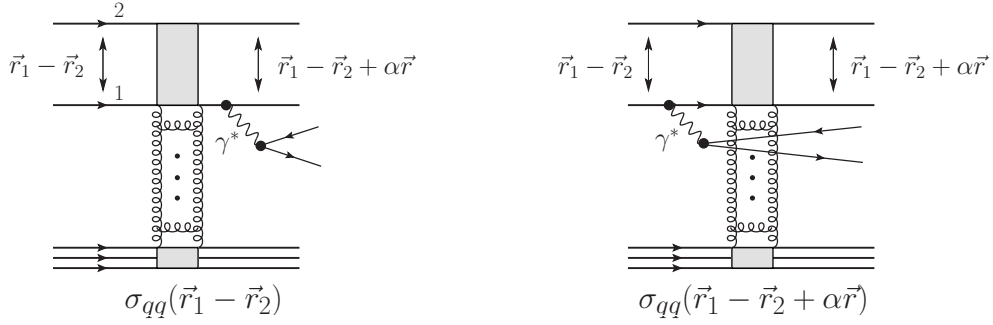,width=13cm}} \caption{Leading
order contribution to the diffractive Drell-Yan in the dipole-target
collision.} \label{fig:gam}
\end{figure}

The hard part of the Drell-Yan process is given by the inelastic
amplitude of $\gamma^*$ radiation by a projectile quark (valence or
sea) due to its interaction with the target through a gluon exchange
as shown in Fig.~\ref{fig:gam}. It consists of two terms
corresponding to interaction of two different Fock states with the
target -- a bare quark before the photon emission $|q\rangle$
($s$-channel diagram), and a quark accompanied by a
Weiz\"acker-Williams photon $|q\gamma^*\rangle$ ($u$-channel
diagram).
%-------------------------------------------------------------
%-------------------------------------------------------------

Elastic dipole scattering depicted in Fig.~\ref{fig:gam} corresponds
to forward scattering at small momentum transfers in the
$t$-channel\footnote{Generally speaking, $\sqrt{-t}\to
\Lambda_{QCD}$ corresponds to the physical forward scattering limit
since transverse momentum of a proton in the final state cannot be
resolved to a better accuracy than its inverse size.}. In the
leading order, the elastic scattering amplitude is given by one-loop
diagram with two $t$-channel gluon exchanges. For on-shell
intermediate spectators, corresponding four-dimensional loop
integral can be reduced to two-dimensional one over the transverse
momentum of one of the gluons
\begin{eqnarray}
2i\mathrm{Im}\,F_{el}(\vec{\delta}_{\perp})=\int\frac{d^2k_{\perp}}{(2\pi)^2}\,
A(\vec{k}_{\perp})A(\vec{\delta}_{\perp}-\vec{k}_{\perp}),\qquad
\vec{\delta}_{\perp}\ll |\vec{k}_{\perp}|
\end{eqnarray}
where $A$ represents (inelastic) amplitude for one $t$-channel gluon
exchange, and the last strong inequality guarantees that the proton
target survives the scattering, hence, the elastic nature of the
process. Then the convolution theorem of Fourier analysis leads to
the optical theorem
\begin{eqnarray}
\mathrm{Im}\,F_{el}(\vec{\delta})=\int
d^2b\,e^{-i\vec{\delta}_{\perp}\cdot\vec{b}}\,\mathrm{Im}\,f_{el}(\vec{b}),\qquad
2i\mathrm{Im}\,f_{el}(\vec{b})=|\tilde{A}(\vec{b})|^2
\label{theorem}
\end{eqnarray}
In the case of multiple elastic rescattering, this relation leads to
eikonalization of the bare elastic amplitude which then correctly
takes into account gap survival effects. However, if one uses the
elastic amplitude $f_{el}$ fitted to soft data, it must already
contain the soft unitarity corrections, and one should include them
twice as is sometimes done in the literature.

Straightforward calculations lead to $\bar qq$ dipole scattering
amplitudes for $s$ and $u$-channel photon emission, respectively,
\begin{eqnarray*}
M^{(1)s}_{\bar
qq}(\vec{b},\vec{r}_p,\vec{r},\alpha)&=&-2ip_i^0\,\sqrt{4\pi}\,\frac{\sqrt{1-\alpha}}{\alpha^2}\,
\Psi^{\mu}_{\gamma^*q}(\alpha,\vec{r})\\
&&\times\frac{1}{N_c}\sum_X\sum_{c_fc_i}\left(\big|V_q(\vec{b})-V_q(\vec{b}+\vec{r_p})\big|^2-\big|V_q(\vec{b}+\vec{r_p})\big|^2\right),\\
M^{(1)u}_{\bar
qq}(\vec{b},\vec{r}_p,\vec{r},\alpha)&=&2ip_i^0\,\sqrt{4\pi}\,\frac{\sqrt{1-\alpha}}{\alpha^2}\,
\Psi^{\mu}_{\gamma^*q}(\alpha,\vec{r})\\
&&\times\frac{1}{N_c}\sum_X\sum_{c_fc_i}
\left(\big|V_q(\vec{b})-V_q(\vec{b}+\vec{r_p}+\alpha\vec{r})\big|^2-\big|V_q(\vec{b}+\vec{r_p}+\alpha\vec{r})\big|^2\right),
\end{eqnarray*}
where the last terms subtract the contributions from diagrams
corresponding to the situation when none of the gluons couple to the
same quark line with the hard photon. Then, implied the fact that
all fields disappear at infinite separations, i.e.
$V_q(\vec{b})\to0$ when $|\vec{b}|\to\infty$, we have due to
antisymmetry of the integrand
\begin{eqnarray}\label{extra-terms}
\int d^2b\,e^{-i\vec{\delta}_{\perp}\cdot\vec{b}}
\Big[\big|V_q(\vec{b}+\vec{r_p})\big|^{2n}-\big|V_q(\vec{b}+\vec{r_p}+\alpha\vec{r})\big|^{2n}\Big]\to0,\quad
n\geq 1\,,\quad |\vec{\delta}_{\perp}|\to0\,,
\end{eqnarray}
such that these terms do not contribute to the final result. Using
the optical theorem for the elastic amplitude
\begin{eqnarray*}
2i\,\mathrm{Im}\,
f_{el}(\vec{b},\vec{r}_p)=\frac{i}{N_c}\sum_X\sum_{c_fc_i}\,\big|V_q(\vec{b})-V_q(\vec{b}+\vec{r_p})\big|^2,
\end{eqnarray*}
we can finally write
\begin{eqnarray}
M^{(1)}_{\bar
qq}(\vec{b},\vec{r}_p,\vec{r},\alpha)=-2ip_i^0\,\sqrt{4\pi}\,\frac{\sqrt{1-\alpha}}{\alpha^2}\,
\Psi^{\mu}_{\gamma^*q}(\alpha,\vec{r})\left[2\mathrm{Im}\,
f_{el}(\vec{b},\vec{r_p})-2\mathrm{Im}\,
f_{el}(\vec{b},\vec{r}_p+\alpha\vec{r})\right]\label{amp-LO}
\end{eqnarray}
i.e. the amplitude of the diffractive radiation is proportional to
the difference between elastic amplitudes of the two Fock
components, with and without the photon radiation. When a quark
fluctuates into the upper Fock quark-photon state with the
transverse separation $\vec{r}$, the final quark gets a transverse
shift $\Delta\vec{r}=\alpha\vec{r}$. Then the quark dipoles with
different sizes in the $|2q\rangle$ and $|2q\gamma^*\rangle$
components interact differently, and their difference corresponds to
the diffractive Drell-Yan amplitude (\ref{amp-LO}).

\section{Diffractive Drell-Yan in proton-proton scattering}

In the dipole picture, the typical enhanced Regge graphs correspond
to elastic scattering of higher Fock states, which contain gluons,
e.g. $|qqg\rangle$, $|qqgg\rangle$, etc. Note that in our approach
we take into account the lowest Fock state $|qq\rangle$ contribution
only. Such an approximation is justified for not very small fraction
$x_{\gamma1}=p_{\gamma}^+/p_1^+$ and scale $M^2$, where valence/sea
quarks are dominated and the gluon contribution is rather small.

The total hadronic amplitude of the diffractive Drell-Yan process
can be written as \cite{KPST06}
\begin{eqnarray}
A_{if}=A^{(1)}_{if}+A^{(2)}_{if}+A^{(3)}_{if}\,,
\end{eqnarray}
where each term corresponds to $\gamma^*$ radiation by one of the
valence (sea) quarks in the proton, in particular,
\begin{eqnarray}
A^{(1)}_{if}(x_{\gamma1},\vec{q}_{\perp},\lambda_{\gamma})&=&\frac{i}{4}\,\alpha^2\int
d^2r_1d^2r_2d^2r_3d^2rd^2bdx_{q_1}dx_{q_2}dx_{q_3}\nonumber\\
&\times&\Psi_{i}(\vec{r}_1,
\vec{r}_2,\vec{r}_3;x_{q_1},x_{q_2},x_{q_3})
\Psi_{f}^*(\vec{r}_1+\alpha
\vec{r},\vec{r}_2,\vec{r}_3;x_{q_1}-x_{\gamma1},x_{q_2},x_{q_3})\nonumber\\
&\times&\Big[M^{\lambda_{\gamma}}_{\bar
qq}(\vec{b},\vec{r}_1-\vec{r}_2,\vec{r},\alpha)+
M^{\lambda_{\gamma}}_{\bar
qq}(\vec{b},\vec{r}_1-\vec{r}_3,\vec{r},\alpha)\Big]
e^{i\vec{l}_{\perp}\cdot\alpha\vec{r}}e^{i\vec{\delta}_{\perp}\cdot\vec{b}}
\label{amp-DDY}
\end{eqnarray}
Here, $\lambda_{\gamma}=L,T$;
$\vec{l}_{\perp}=\vec{\delta}_{\perp}-\vec{q}_{\perp}/\alpha$
($z$-axis is directed along initial proton momentum); the hard
photon with virtuality $q^2=M^2\gg m_p^2$, transverse
$\vec{q}_{\perp}$ and fractional longitudinal $x_{\gamma1}$ momenta
is emitted from the first valence quark with impact parameter
$\vec{r}_1$, other two valence quarks in the proton have impact
parameters $\vec{r}_2$ and $\vec{r}_3$, respectively; $\vec{r}$ is
transverse separation between the photon and the radiating quark;
$\alpha=x_{\gamma1}/x_{q_1}$ is the fraction of longitudinal momenta
taken away by the photon from the radiating quark; $M^{L,T}_{\bar
qq}$ are the Fourier-transformed amplitudes for the elastic quark
dipole scattering off the proton target accompanied by the hard
$L,T$-polarized photon emission; $\Psi_{i,f}$ are the light-cone
wave functions of the $3q$ systems in the initial and final state,
respectively. In Eq.~(\ref{amp-DDY}) we implicitly assumed that
exchanges $t$-channel gluons all together take a negligibly small
longitudinal momentum compared to the collisions energy $\sqrt{s}$
and, hence, corrections to quark momenta due to gluon couplings are
neglected in the wave functions.

The differential cross section for the single diffractive di-lepton
production in the target rest frame reads
\begin{eqnarray}\nonumber
d^{\,8}\sigma^{sd}_{\lambda_{\gamma}}(pp\to
pl^+l^-X)&=&\sum_f\sum_{n=1}^3
|A^{(n)}_{if}(x_{\gamma1},\vec{q}_{\perp},\lambda_{\gamma})|^2\frac{d\alpha}{\alpha(1-\alpha)}\frac{
d^2q_{\perp}d^2\delta_{\perp}}{(2\pi)^5\,8(p_{i,n}^0)^2}
\\
&\times&\alpha_{em}\epsilon_{\mu}(\lambda_{\gamma})\epsilon^*_{\nu}(\lambda_{\gamma})L^{\mu\nu}
\frac{dM^2d\Omega}{16\pi^2M^4},\quad \lambda_{\gamma}=L,T
\label{cross-sec}
\end{eqnarray}
where prefactors provide averaging over colors and helicities of
exchanged $t$-channel gluons, $p_{i,n}^0$ is the energy of the
radiating $n$th quark in the initial state, $n=1,\,...,\,3$;
$\alpha_{em}=e^2/(4\pi)=1/137$ is the electromagnetic coupling
constant. The second line in Eq.(\ref{cross-sec}) describes decay of
$\gamma^*$ into the leptonic pair $l^+l^-$ into solid angle
$d\Omega=d\phi d\cos\theta$, and $L^{\mu\nu}$ is the standard
leptonic tensor. We keep in the cross section only diagonal in the
photon polarization $\lambda_{\gamma}=L,T$ terms (non-diagonal ones
drop out after integration over leptonic azimuthal angle $\phi$).
Integrating the diffractive differential DY cross section over the
photon transverse momentum $\vec{q}_{\perp}$ we get
\begin{eqnarray}\label{DDY-cs}
\frac{d^4\sigma_{L,T}(pp\to
pl^+l^-X)}{d\ln\alpha\,dM^2\,d^2\delta_{\perp}}=
\frac{\alpha_{em}}{3\pi M^2}\,\frac{d^3\sigma_{L,T}(pp\to
p\gamma^*X)}{d\ln\alpha\,d^2\delta_{\perp}}\,.
\end{eqnarray}
Then applying the completeness relation
\begin{eqnarray}\nonumber
&&\sum_f\Psi_f(\vec{r}_1+\alpha
\vec{r},\vec{r}_2,\vec{r}_3;x_{q_1},x_{q_2},x_{q_3})
\Psi^*_f(\vec{r}\,'_1+\alpha
\vec{r}\,',\vec{r}\,'_2,\vec{r}\,'_3;x'_{q_1},x'_{q_2},x'_{q_3})\\
&&\phantom{.......}=\,
\delta(\vec{r}_1-\vec{r}\,'_1+\alpha(\vec{r}-\vec{r}\,'))\delta(\vec{r}_2-\vec{r}\,'_2)
\delta(\vec{r}_3-\vec{r}\,'_3)\prod_{j=1}^3\delta(x_{q_j}-x'_{q_j})
\end{eqnarray}
we get the diffractive $\gamma^*$ production cross section in the
following differential form
\begin{eqnarray}
&&\frac{d^3\sigma_{\lambda_{\gamma}}(pp\to
p\gamma^*X)}{d\ln\alpha\,d^2\delta_{\perp}}=\frac{\sum_q
Z_q^2}{64\pi^2}\int
d^2r_1d^2r_2d^2r_3d^2r\,d^2bd^2b'\,dx_{q_1}dx_{q_2}dx_{q_3}\nonumber\\
&&\qquad\qquad\times\,|\tilde{\Psi}^{\lambda_{\gamma}}_{\gamma^*q}(\alpha,\vec{r})|^2|\Psi_{i}(\vec{r}_1,
\vec{r}_2,\vec{r}_3;x_{q_1},x_{q_2},x_{q_3})|^2\nonumber\\
&&\qquad\qquad\times\,\Delta(\vec{r}_1,
\vec{r}_2,\vec{r}_3;\vec{b};\vec{r},\alpha)\Delta(\vec{r}_1,
\vec{r}_2,\vec{r}_3;\vec{b}\,';\vec{r},\alpha)\,
e^{i\vec{\delta}_{\perp}\cdot(\vec{b}-\vec{b}\,')} \label{eik-tot}
\end{eqnarray}
where $\tilde{\Psi}_{\gamma^*q}=\Psi_{\gamma^*q}/Z_q$, and
\begin{eqnarray}\nonumber
\Delta&=&-2\mathrm{Im}\,
f^{\mathrm{KST}}_{el}(\vec{b},\vec{r}_1-\vec{r}_2)+2\mathrm{Im}\,
f^{\mathrm{KST}}_{el}(\vec{b},\vec{r}_1-\vec{r}_2+\alpha\vec{r})\\&&-2\mathrm{Im}\,
f^{\mathrm{KST}}_{el}(\vec{b},\vec{r}_1-\vec{r}_3)+2\mathrm{Im}\,
f^{\mathrm{KST}}_{el}(\vec{b},\vec{r}_1-\vec{r}_3+\alpha\vec{r})\,,
\label{eik-el}
\end{eqnarray}
where the Kopeliovich-Sch\"afer-Tarasov (KST) parameterization of
the elastic dipole-target amplitude fitted to the soft data
\cite{KST-GBW-eqs,kpss,KST-par} and, hence, valid at
$|\vec{r}_i-\vec{r}_j|\sim \vec{b},\,i\not=j$ is used. Finally,
going over to the forward limit $\vec{\delta}_{\perp}=0$ we obtain
\begin{eqnarray}
&&\frac{d^3\sigma_{\lambda_{\gamma}}(pp\to
p\gamma^*X)}{d\ln\alpha\,d\delta_{\perp}^2}\Big|_{\delta_{\perp}=0}=\frac{\sum_q
Z_q^2}{64\pi}\int
d^2r_1d^2r_2d^2r_3d^2r\,dx_{q_1}dx_{q_2}dx_{q_3}\nonumber\\
&&\qquad\times\,|\tilde{\Psi}^{\lambda_{\gamma}}_{\gamma^*q}
(\alpha,\vec{r})|^2|\Psi_{i}(\vec{r}_1,
\vec{r}_2,\vec{r}_3;x_{q_1},x_{q_2},x_{q_3})|^2\,\left[\int d^2b\,
\Delta(\vec{r}_1,
\vec{r}_2,\vec{r}_3;\vec{b};\vec{r},\alpha)\right]^2
\end{eqnarray}
We see that normalization of the cross section agrees with the
original result of Ref.~\cite{KPST06}. The total diffractive cross
section is then given by
\begin{eqnarray}
\frac{d\sigma(pp\to
p\gamma^*X)}{d\ln\alpha}=\frac{1}{B_{sd}^{DY}(s)}\frac{d^3\sigma(pp\to
p\gamma^*X)}{d\ln\alpha\,d\delta_{\perp}^2}\Big|_{\delta_{\perp}=0}
\end{eqnarray}
where $B_{sd}^{DY}(s)$ is the diffractive slope similar to the one
measured in diffractive DIS.

The next step is to introduce the proton wave function assuming the
Gaussian shape for the quark distributions in the proton as
\begin{eqnarray}\nonumber
|\Psi_i(\vec{r}_1,
\vec{r}_2,\vec{r}_3;x_{q_1},x_{q_2},x_{q_3})|^2&=&\frac{2+a/b}{\pi^2ab}
\exp\Big[-\frac{r_1^2}{a}-\frac{r_2^2+r_3^2}{b}\Big]\rho(x_{q_1},x_{q_2},x_{q_3})\\
&\times&\delta(\vec{r}_1+\vec{r}_2+\vec{r}_3)\delta(1-x_{q_1}-x_{q_2}-x_{q_3})
\label{psi}
\end{eqnarray}
where $a=\langle r_{\bar qq}^2 \rangle$ and $b=\langle R_q^2
\rangle$ are the diquark mean radius squared and the quark mean
distance from the diquark squared, respectively. In this work, we
will use the simplest case of symmetric valence quarks distribution
assuming that $r_{\bar qq}=R_q=0.85$ fm.

Then valence quark distribution in the proton is given by
\begin{eqnarray}\nonumber
\int
dx_{q_2}dx_{q_3}\rho(x_{q_1},x_{q_2},x_{q_3})=\rho_{q_1}(x_{q_1})\,.
\end{eqnarray}
where we integrated out the longitudinal fractions of the diquark in
the proton. Generalization of the three-body proton wave function
(\ref{psi}) including different quark and antiquark flavors, as well
as sea quarks, leads to the proton structure function \cite{KRT00}
\begin{eqnarray}\label{SF}
\sum_q Z_q^2[\rho_q(x_q)+\rho_{\bar
q}(x_q)]=\frac{1}{x_q}F_2(x_q),\qquad
x_q=\frac{x_{\gamma1}}{\alpha}\,.
\end{eqnarray}

\section{Numerical results for differential DDY cross sections}

\begin{figure}[!h]
\begin{minipage}{0.49\textwidth}
 \centerline{\includegraphics[width=1.0\textwidth]{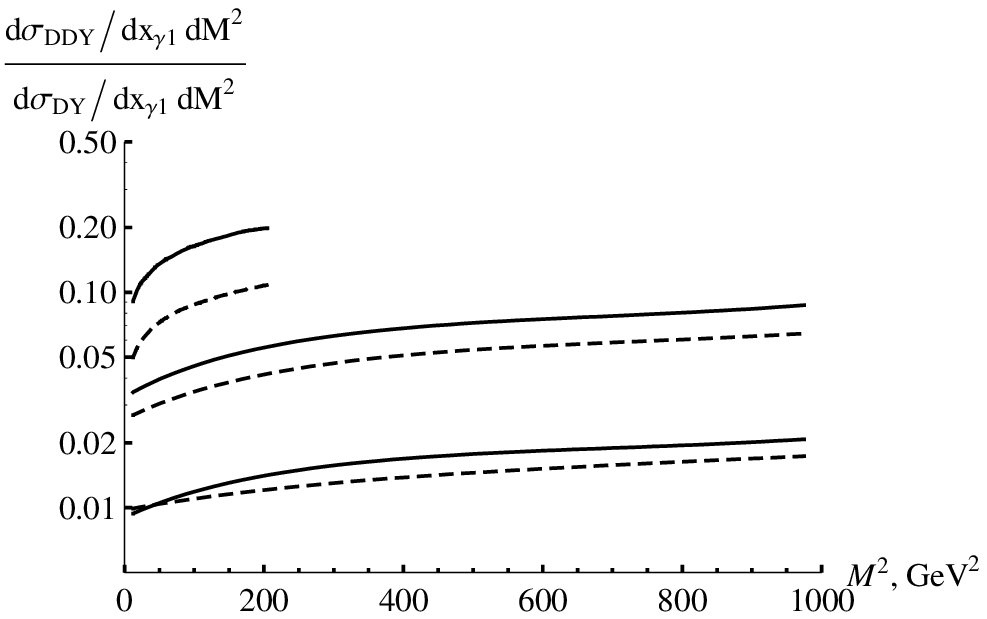}}
\end{minipage}
\hspace{0.5cm}
\begin{minipage}{0.46\textwidth}
 \centerline{\includegraphics[width=1.0\textwidth]{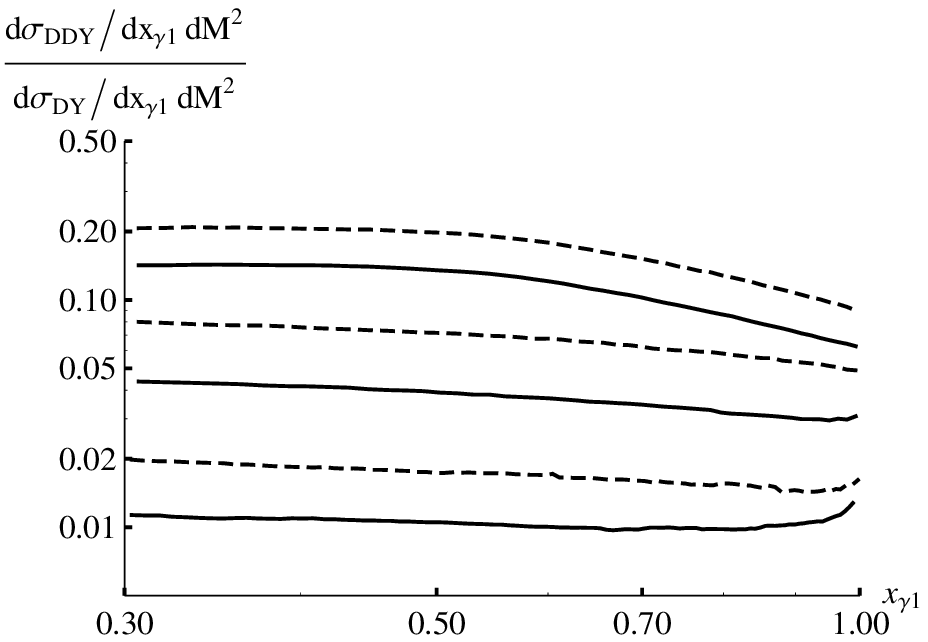}}
\end{minipage}
   \caption{
\small The ratio of the diffractive to inclusive Drell-Yan cross
sections as function of the lepton-pair invariant mass squared $M^2$
(left panel) and photon fraction $x_{\gamma1}$ (right panel) at
different energies.}
 \label{fig:DDYvsDY}
\end{figure}

In Fig.~\ref{fig:DDYvsDY} the ratio of the diffractive to inclusive
DY cross sections is plotted as a function of di-lepton invariant
mass squared $M^2$ (left panel) and photon fractional light-cone
momentum $x_{\gamma1}$ (right panel) at different energies. In the
left panel, the curves are given for fixed $x_{\gamma1}=0.5$ (solid
lines) and $x_{\gamma1}=0.9$ (dashed lines). In the right panel, the
curves are given for fixed $M^2=50\,\GeV^2$ (solid lines) and
$M^2=500\,\GeV^2$ for $\sqrt{s}=14$ TeV and 500 GeV, and
$M^2=200\,\GeV^2$ at $\sqrt{s}=40$ GeV (dashed lines). The pairs of
solid/dashed curves in the both panels correspond to
$\sqrt{s}=40\,\GeV$, $500$ GeV and $14$ TeV from top to bottom,
respectively. Here we used the KST parameterization for the
dipole-target scattering amplitude \cite{kpss,KST-GBW-eqs,KST-par}
and $F_2$ parameterization by Cudell and Soyez \cite{Cudell-F2} are
used here. In this calculation we consider the unpolarized case
summing up the contributions of longitudinal and transverse parts
both in the diffractive and inclusive cross sections.
%-----------------------------------------------------------------------------
%------------------------------------------------------------------------------

As seen from Fig.~\ref{fig:DDYvsDY}, the DDY-to-DY cross section
ratio is falling with energy. However, naively one could expect
basing on QCD factorisation, that the DDY cross section, which is
proportional to the dipole cross section squared, should rise with
energy steeper than the total inclusive cross section. At the same
time, the ratio rises with the hard scale of the process, $M^2$.
This also looks counterintuitive, since diffraction is usually
associated with soft interactions \cite{deriv1}. These effects are
different from ones emerging in Regge factorisation-based
calculations, where we observe a slow rise of the DDY-to-DY cross
section ratio with c.m.s. energy and its practical independence on
the hard scale of the process $M^2$ \cite{Antoni11}.

In order to understand such an interesting behavior of the DDY-to-DY
cross sections ratio obtained in the color dipole approach, let us
look at the amplitude of the DDY process, which is proportional to
the difference between the dipole cross sections of the Fock states
with and without the hard photon emission \cite{KPST06}, i.e.
\begin{eqnarray}
M_{DDY}\sim\sigma(\vec{R})-\sigma(\vec{R}-\alpha\vec{r})=
\frac{2\alpha\sigma_0}{R_0^2(x)}e^{R^2/R_0^2(x)}\,
\left(\vec{r}\cdot\vec{R}\right)+h.o.
\end{eqnarray}
assuming the simplest Golec-Biernat-Wusthoff (GBW) slope for the
dipole cross section \cite{GBWdip}, and the hardness of the emitted
photon implies $r\sim 1/M\ll R_0(x)$. We see now that the
diffractive DY amplitude is linear in $r$, so the diffractive cross
section turns out to be a quadratic function of $r$, which is
different from e.g. the diffractive DIS process where the cross
section is proportional to $r^4$ and is dominated by soft
fluctuations (see e.g. Refs.~\cite{KPSdiff,BP97}). Since the
diffractive DY cross section is proportional to $r^2$, then soft and
hard interactions contribute on the same footing \cite{KPST06},
which is one of the basic sources of the QCD factorisation breaking
in diffractive DY process.

%-----------------------------------------------------------------------------
\begin{figure}[!h]
\begin{minipage}{0.49\textwidth}
 \centerline{\includegraphics[width=1.0\textwidth]{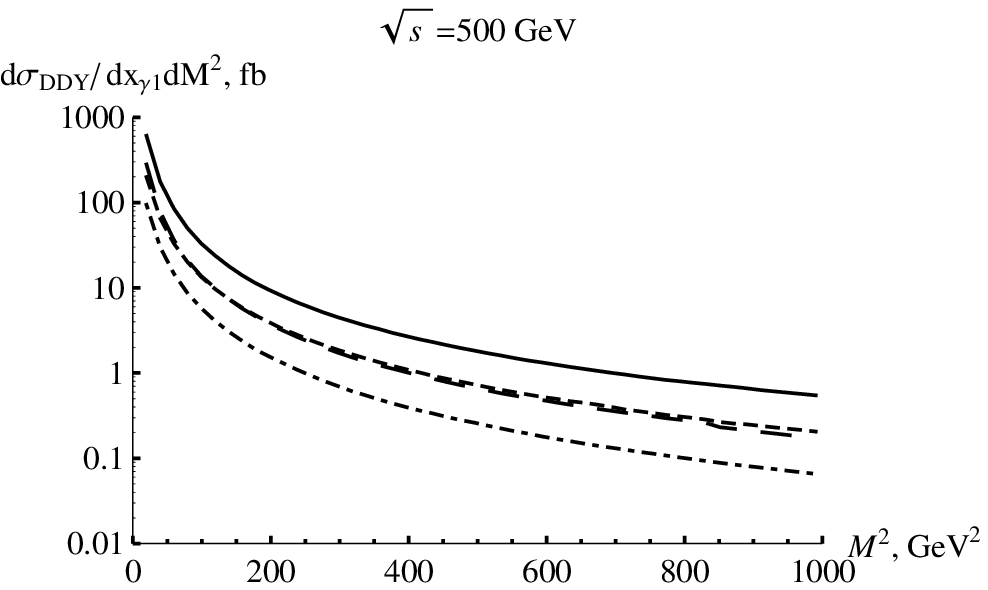}}
\end{minipage}
\hspace{0.5cm}
\begin{minipage}{0.46\textwidth}
 \centerline{\includegraphics[width=1.0\textwidth]{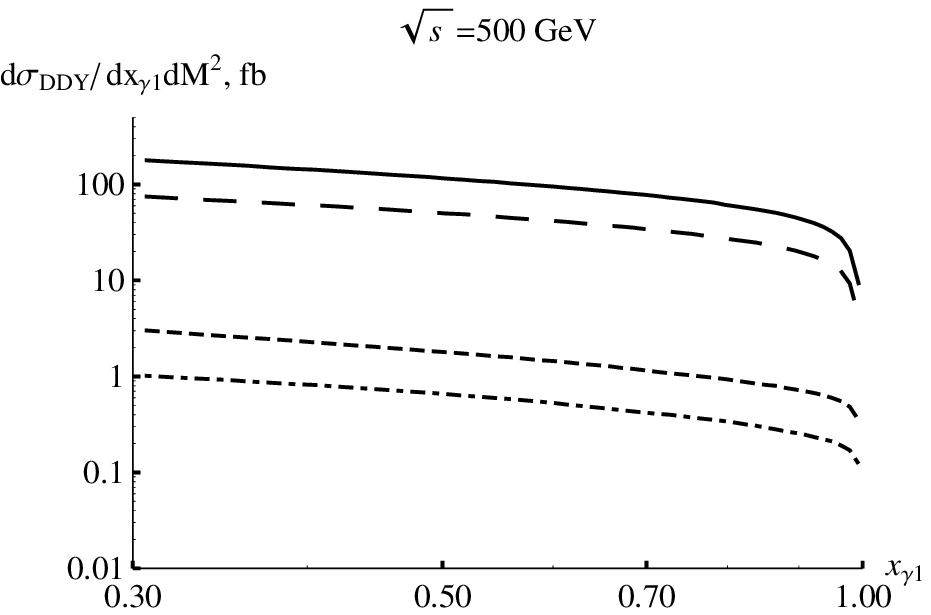}}
\end{minipage}
   \caption{
\small Diffractive Drell-Yan cross section (in fb) as function of
the lepton-pair invariant mass squared $M^2$ (left panel) and photon
fraction $x_{\gamma1}$ (right panel) at the RHIC II c.m.s. energy
$\sqrt{s}=500$ GeV.}
 \label{fig:GBWvsKST}
\end{figure}
%------------------------------------------------------------------------------
We also compare predictions for the diffractive DY cross section for
different parameterizations for elastic dipole-target scattering
amplitude corresponding to scattering of small (GBW given by
Refs.~\cite{GBWdip,KST-GBW-eqs,GBW-par}) and large (KST given by
Refs.~\cite{KST-GBW-eqs,kpss,KST-par}) dipoles. As an example, in
Fig.~\ref{fig:GBWvsKST} we present the diffractive Drell-Yan cross
section as function of the lepton-pair invariant mass squared $M^2$
(left panel) and photon fraction $x_{\gamma1}$ (right panel) at the
RHIC II c.m.s. energy $\sqrt{s}=500$ GeV. We notice that the GBW
parameterization leads to roughly a factor of two smaller cross
section than the one obtained with the KST parameterization,
however, both of them exhibit basically the same $x_{\gamma1}$ and
$M^2$ shapes. It means that the evolution of the dipole size can
only affect the overall normalization of the DDY cross section.
Since arguments in the elastic amplitude $f_{el}({\vec r}_p,{\vec
b})$, the impact distance between the target and the projectile $b$
and the transverse distance between projectile quarks
$r_p\sim|\vec{r}_i-\vec{r}_j|,\,i\not=j$, are of the same order and
given at the soft hadronic scale, then the use of KST
parameterization fitted to the soft hadron scattering data data is
justified in the case of diffractive DY.

\section{Conclusion}

The QCD factorisation breaking effects in the diffractive Drell-Yan
process lead to quite different properties of the corresponding
observables with respect to QCD factorisation-based calculations. A
quark cannot diffractively radiate a photon in the forward
direction, whereas a hadron can due to the presence of transverse
motion of spectator quarks in the projectile hadron. For this
reason, the diffractive DY cross section depends on the hadronic
size explicitly breaking the QCD factorisation.

This leads to the physical picture where hard and soft interactions
are equally important for DY diffraction, and their relative
contributions are independent of the hard scale, like in the
inclusive DY process. This is a result of the specific property of
DY diffraction: its cross section is a linear, rather than quadratic
function of the dipole cross section. On the contrary, diffractive
DIS is predominantly a soft process, because its cross section is
proportional to the dipole cross section squared.

Contrary to what follows from the calculations based on QCD
factorisation, the ratio of the diffractive to inclusive cross
sections falls with energy, but rises with the di-lepton effective
mass $M$. This happens due to the saturated behavior of the dipole
cross section which levels off at large separations. All these
properties are different from those in the diffractive DIS process,
where QCD factorisation is exact. In addition, we made predictions
for the differential (in photon fractional momentum $x_{\gamma1}$
and di-lepton invariant mass squared $M^2$) cross sections for the
diffractive DY process at the energies of RHIC (500 GeV) and LHC (14
TeV).

\subsection{Acknowledgments}

Useful discussions and helpful correspondence with Jochen Bartels,
Antoni Szczurek, Gunnar Ingelman and Mark Strikman are gratefully
acknowledged. This study was partially supported by the Carl Trygger
Foundation (Sweden), by Fondecyt (Chile) grant 1090291, and by
Conicyt-DFG grant No. 084-2009. Authors are also indebted to the
Galileo Galilei Institute of Theoretical Physics (Florence, Italy)
and to the INFN for partial support and warm hospitality during
completion of this work.

\section*{References}
%===========================


\begin{thebibliography}{99}
%===========================

\bibitem{KPST06}
  B.~Z.~Kopeliovich, I.~K.~Potashnikova, I.~Schmidt, A.~V.~Tarasov,
  %``Unusual features of Drell-Yan diffraction,''
  Phys.\ Rev.\  {\bf D74}, 114024 (2006)
  [hep-ph/0605157].


\bibitem{heavyF}
  B.~Z.~Kopeliovich, I.~K.~Potashnikova, I.~Schmidt, A.~V.~Tarasov,
  %``Diffractive Excitation of Heavy Flavors: Leading Twist Mechanisms,''
  Phys.\ Rev.\  {\bf D76}, 034019 (2007)
  [hep-ph/0702106].

\bibitem{Glauber}
R.~J.~Glauber, Phys. Rev. 100, 242 (1955).

\bibitem{FP56}
E.~Feinberg and I.~Ya.~Pomeranchuk, Nuovo. Cimento. Suppl. 3 (1956)
652.

\bibitem{GW}
M.~L.~Good and W.~D.~Walker, Phys. Rev. 120 (1960) 1857.

\bibitem{KPSdiff}
  B.~Z.~Kopeliovich, I.~K.~Potashnikova, I.~Schmidt,
  %``Diffraction in QCD,''
  Braz.\ J.\ Phys.\  {\bf 37}, 473-483 (2007).
  [arXiv:hep-ph/0604097 [hep-ph]].

\bibitem{BP97}
B.~Z.~Kopeliovich and B.~Povh, Z. Phys. {\bf A356}, 467 (1997).

\bibitem{PCAC}
  B.~Z.~Kopeliovich, I.~K.~Potashnikova, I.~Schmidt, M.~Siddikov,
  %``Breakdown of PCAC in diffractive neutrino interactions,''
  Phys.\ Rev.\  {\bf C84}, 024608 (2011)
  [arXiv:1105.1711 [hep-ph]].

\bibitem{KST-GBW-eqs}
B.~Z.~Kopeliovich, A.~H.~Rezaeian, I.~Schmidt,
%``Azimuthal Asymmetry of pions in pp and pA collisions,''
Phys.\ Rev.\  {\bf D78}, 114009 (2008) [arXiv:0809.4327 [hep-ph]].

\bibitem{kpss}
  B.~Z.~Kopeliovich, I.~K.~Potashnikova, I.~Schmidt and J.~Soffer,
  %``Damping of forward neutrons in $p p$ collisions,''
  Phys.\ Rev.\  D {\bf 78}, 014031 (2008)
  [arXiv:0805.4534 [hep-ph]].

\bibitem{KST-par}
B.~Z.~Kopeliovich, A.~Sch\"afer and A.~V.~Tarasov, Phys. Rev. {\bf
D62}, 054022 (2000) [arXiv:hep-ph/9908245];\\
B.~Z.~Kopeliovich, I.~K.~Potashnikova, I.~Schmidt, J.~Soffer,
%``Damping of forward neutrons in $p p$ collisions,''
Phys.\ Rev.\  {\bf D78}, 014031 (2008) [arXiv:0805.4534 [hep-ph]].

\bibitem{KRT00}
  B.~Z.~Kopeliovich, J.~Raufeisen, A.~V.~Tarasov,
  %``The Color dipole picture of the Drell-Yan process,''
  Phys.\ Lett.\  {\bf B503}, 91-98 (2001).
  [hep-ph/0012035].

\bibitem{Cudell-F2}
  J.~R.~Cudell, G.~Soyez,
  %``Does F(2) need a hard pomeron?,''
  Phys.\ Lett.\  {\bf B516}, 77-84 (2001).
  [hep-ph/0106307].

\bibitem{deriv1}
B. Z. Kopeliovich, proc. of the workshop Hirschegg '95: Dynamical
Properties of Hadrons in Nuclear Matter, Hirschegg January 16-21,
1995, ed. by H. Feldmeyer and W. N\"orenberg, Darmstadt, 1995, p.
102 [hep-ph/9609385].

\bibitem{Antoni11}
  G.~Kubasiak, A.~Szczurek,
  %``Inclusive and exclusive diffractive production of dilepton pairs in proton-proton collisions at high energies,''
  Phys.\ Rev.\  {\bf D84}, 014005 (2011).
  [arXiv:1103.6230 [hep-ph]].

\bibitem{GBWdip}
  K.~J.~Golec-Biernat, M.~Wusthoff,
  %``Saturation effects in deep inelastic scattering at low Q**2 and its implications on diffraction,''
  Phys.\ Rev.\  {\bf D59}, 014017 (1998).
  [hep-ph/9807513].

\bibitem{GBW-par}
B.~Z.~Kopeliovich, H.~J.~Pirner, A.~H.~Rezaeian and I.~Schmidt,
Phys. Rev. {\bf D77}, 034011 (2008) [arXiv:0711.3010 [hep-ph]].


\end{thebibliography}
\end{document}